\documentclass[aps,prl,superscriptaddress,twocolumn,showpacs]{revtex4}

\newcommand{\SinEff}{$\sin^2\theta^{\mbox{\footnotesize lept}}_{\mbox{\footnotesize eff}}$}
\newcommand{\sineff}{\sin^2\theta^{\mbox{\footnotesize lept}}_{\mbox{\footnotesize eff}}}

\usepackage{epsfig}

\begin{document}

\title{Complete Two-Loop Electroweak Fermionic Corrections to
sin$^2\theta^{\mbox{\small lept}}_{\mbox{\small eff}}$\\ and 
Indirect Determination of the Higgs Boson Mass}

\author{M. Awramik}

\affiliation{DESY, Platanenallee 6, D-15738 Zeuthen, Germany}

\affiliation{Institute of Nuclear Physics PAS, Radzikowskiego 152,
  PL-31342 Cracow, Poland}

\author{M. Czakon}

\affiliation{DESY, Platanenallee 6, D-15738 Zeuthen, Germany}

\affiliation{Institute of Physics, University of Silesia, Uniwersytecka 4,
  PL-40007 Katowice, Poland}

\author{A. Freitas}

\affiliation{Theoretical Physics Division, Fermilab, P. O. Box 500, Batavia, IL
  60510, USA}

\author{G. Weiglein}

\affiliation{Institute for Particle Physics Phenomenology, University of
Durham, Durham DH1~3LE, UK}

\begin{abstract}

We present a complete calculation of the contributions to the
effective leptonic weak mixing angle, \SinEff, generated by closed
fermion loops at the two-loop level of the electroweak
interactions. This quantity is the source of the most stringent bound
on the mass, $M_H$, of the only undiscovered particle of the Standard
Model, the Higgs boson. The size of the corrections with respect to
known partial results varies between $-4\times 10^{-5}$ and $-8\times
10^{-5}$ for a realistic range of $M_H$ from 100 to 300 GeV. This
translates into a shift of the predicted (from \SinEff alone) central
value of $M_H$ by +19  GeV, to be compared with the shift induced by a
recent change in the measured top quark mass which amounts to +36
GeV. Our result, together with all other known corrections is given in
the form of a precise fitting formula to be used in the global fit to
the electroweak data.

\end{abstract}


\maketitle

The search for the Standard Model Higgs boson lies among the most
important objectives of present elementary particle physics. The
experimental discovery will be possible at the Large Hadron Collider
(LHC) within a mass range reaching up to 1 TeV.  On the other hand, it
is more than desirable to have as stringent indirect bounds on $M_H$
as possible with the help of precision measurements. Should the Higgs
boson be discovered, these bounds will serve as a strong test of the
model.

In this letter we study the quantity that has the highest weight in
the combined fit to electroweak data as far as $M_H$ prediction is
concerned, which is the effective leptonic weak mixing angle,
\SinEff. It can be defined through the form factors at the $Z$
boson  pole of the vertex coupling the $Z$ to leptons ($l$).
If this vertex is written as $i\; \overline{l} \gamma^\mu (g_V-g_A
\gamma_5)l \; Z_\mu$ then
\begin{equation}
  \label{definition}
  \sineff = 1/4
  \left(1-\mbox{Re}\left(g_V/g_A\right)\right).
\end{equation}
At  tree-level this  amounts  to the  sine  of the  weak mixing  angle
$\sin^2 \theta_{\rm  W} = 1-M_W^2/M_Z^2$  in the on-shell  scheme.  In
practice, \SinEff is derived  from various asymmetries measured around
the $Z$  boson peak  at $e^+ e^-$  colliders after subtraction  of QED
effects.  The  current experimental  value  is  $0.23150 \pm  0.00016$
\cite{exp}. The  high precision  quoted and the  expected size  of the
radiative  corrections make  the  result indispensable  for a  precise
prediction of $M_H$. A lot of effort has been put into the theoretical
calculation  of  \SinEff.  Besides  the  one-loop  contributions  also
higher-order  QCD corrections  \cite{QCD2L,QCD3L} are  known. However,
for the electroweak two-loop corrections, only the leading term in the
large   $M_H$  expansion   \cite{vanderBij:1983bw}  and   the  leading
\cite{Fleischer:1993ub} and subleading \cite{Degrassi:1996ps} terms in
the large top  quark mass expansion are available up  to now. The goal
of  the present  work  is  the calculation  of  the complete  two-loop
electroweak contributions with one or two closed fermion loops.

The  prediction Eq.~(\ref{definition})  does  not use  $M_W$ as  input
parameter,  but  the results  are  given  by  using the  very  precise
measurement  of  the Fermi  constant,  $G_\mu$,  from  the muon  decay
lifetime to derive $M_W$.   Consequently the calculation of \SinEff as
a function of  $G_\mu$ involves also the computation  of the radiative
corrections  to  the relation  between  $G_\mu$  and  $M_W$.  For  the
electroweak two-loop corrections  with closed fermion loops considered
here,  this  has  been  carried  out in  Ref.~\cite{muon}.   We  will,
therefore, also use the quantity $\Delta\kappa$,
\begin{equation}
  \sineff = \left(1- M_W^2/M_Z^2 \right)  \left(1+
  \Delta\kappa \right),
  \label{kappa}
\end{equation}
which is only weakly sensitive to $M_W$, but encompasses the loop
corrections to the $Z$ form factors.

\begin{figure}
\hspace{0cm} \epsfig{file=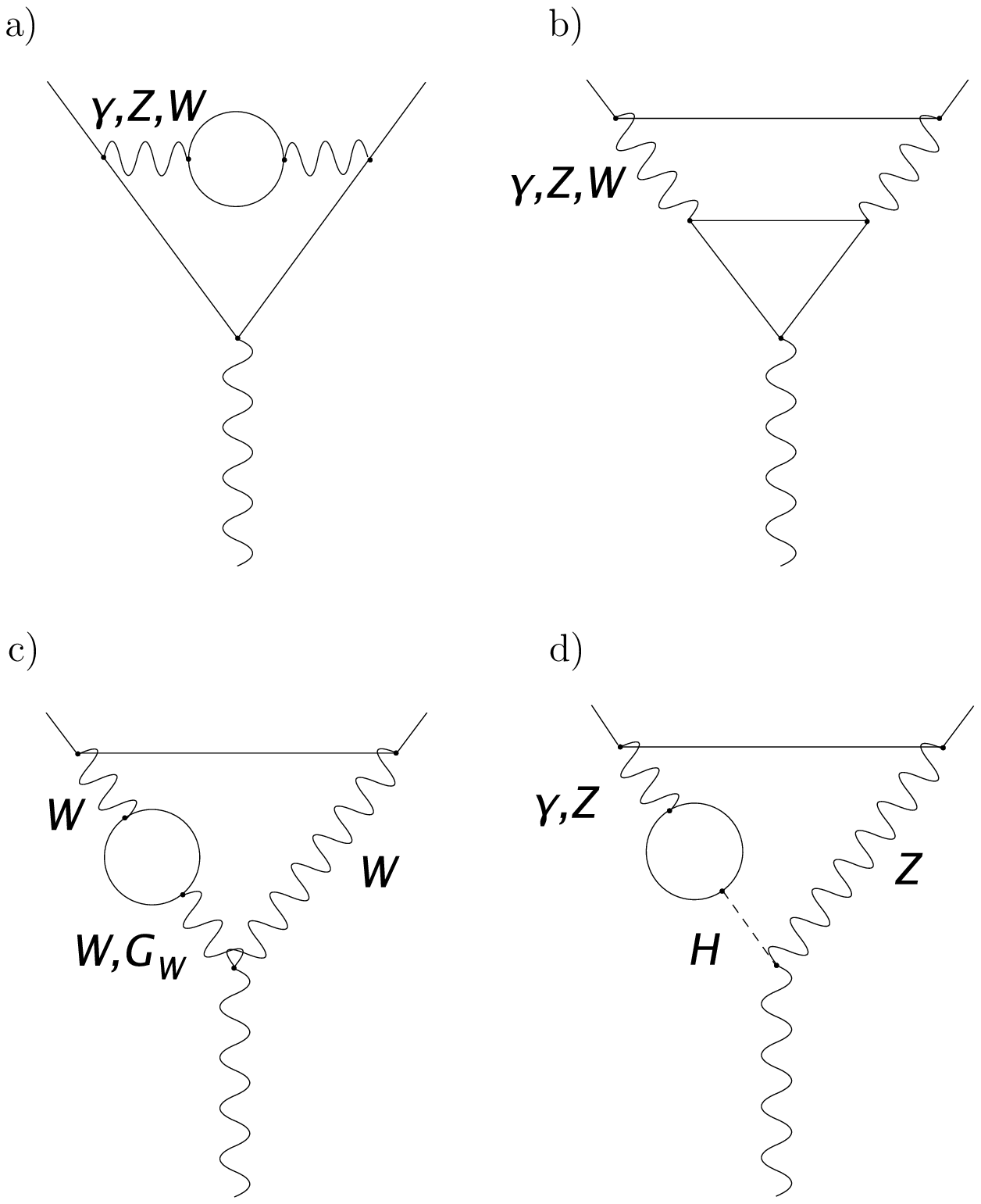,width=5cm}
\caption{\label{diags}
Two-loop vertex diagrams entering the calculation.}
\end{figure}

We focus in this letter on the discussion of our main results. A
detailed description of the calculation will be given in a forthcoming
publication. Here, we note only that the
contributions to the form factors can be divided into two major
parts. The first one comprises the terms from renormalization.  We use
the on-shell renormalization scheme, similarly to the previous
calculation of $M_W$ \cite{muon}.  The second one consists of the bare
two-loop vertex diagrams, the total number of which approaches five
hundred. Upon restriction to those containing a closed fermion loop we
count only a few tens, which can be cast into four topologies as shown
in Fig.~\ref{diags}.  There is no dependence on the Higgs boson mass
in the pure two-loop  vertex diagrams, since {\it CP} conservation
makes d) vanish. It is convenient to subdivide the remaining diagrams
into those containing a top quark line and those containing only light
fermion lines. The former can be evaluated with the large top quark
mass expansion, using the smallness of the ratio $M_Z^2/m_t^2 \approx
1/4$. We have convinced ourselves that an expansion up to
$(M_Z^2/m_t^2)^5$ is sufficient to obtain an intrinsic precision of
the order of $10^{-7}$. The diagrams with only light fermion lines
introduce also an important simplification: they have just two scales
at most, $M_W$ and $M_Z$, since at the level we are considering, light
fermion masses can be safely neglected. The problem thus contains only
one variable, and lends itself naturally to the approach of
differential equations \cite{Kotikov:hm}. A prerequisite for this
method is the complete reduction of the integrals to a small set of
independent masters. This has been achieved with the C++ library
DiaGen/IdSolver \cite{idsolver},  and been checked for a number of
diagrams by an independent calculation. At the end we obtained
analytic expressions for all of the integrals but one. The latter,
corresponding to diagram b) of Fig.~\ref{diags}, has been evaluated by
a one dimensional integral representation.  All integrals have been
checked by different expansions in physical and unphysical regimes and
by numerical integrations based on dispersion relations
\cite{Bauberger:1994by} and Feynman parameterizations
\cite{Ghinculov:1994sd}.

An interesting problem connected to two-loop vertex diagrams is the
treatment of the $\gamma_5$ matrix in triangle fermion loops. We used
the naive dimensional regularization with a four-dimensional treatment
of resulting epsilon tensors as already explained in \cite{muon}. We
observed, however, that the contributions are divergent due to the
soft-collinear behavior of the diagrams with external on-shell
massless fermions. This would undermine the correctness of the
approach if the dimension of space-time were the only regulator. We
decided to use a finite photon mass as the regulator at the expense of
a subsequent difficult expansion corresponding to a mixed
Sudakov/threshold regime. The difference between the full result and
the result which would be obtained if all traces containing a single
$\gamma_5$ were set to zero will be denoted in what follows with a
tr$\gamma_5$ subscript.

We shall now discuss the numerical effect of the new two-loop result
for the effective weak mixing angle.  We focus on the contributions to
$\Delta\kappa$, Eq.~(\ref{kappa}), taking the current experimental
value for $M_W$ as input.  The associated error is not relevant for
the analysis, since the final prediction uses $G_\mu$ as input,
combining the radiative corrections to $M_W$ and $\Delta\kappa$.  We
use the parameter values given in Tab.~\ref{input}.  Note that the
experimentally determined $W$ and $Z$ boson masses correspond to a
Breit-Wigner parametrization with a running width and  have to be
translated to the pole mass scheme used in our calculation
\cite{muon}, resulting in a downward shift \cite{riemann}. For $M_W$
and $M_Z$, this shift amounts to about 27 and 34 MeV, respectively.

\begin{table}
  \caption{\label{input} Input parameters with errors where relevant
  for the present analysis.}
  \begin{ruledtabular}
    \begin{tabular}{ll}
      input parameter & value\\ 
      \hline 
      $M_W$ & $80.426{\rm \; GeV}$ \\
      $M_Z$ & $91.1876 \pm 0.0021{\rm \; GeV}$ \\
      $\Gamma_Z$ & $2.4952 {\rm \; GeV}$ \\
      $m_t$ & $178.0 \pm 4.3 {\rm \; GeV}$ \\
      $m_b$ & $4.85 {\rm \; GeV}$ \\
      $\Delta\alpha(M_Z^2)$ & $0.05907 \pm 0.00036$ \\
      $\alpha_s(M_Z)$ & $0.117 \pm 0.002$ \\
      $G_\mu$ & $1.16637 \times 10^{-5} {\rm \; GeV^{-2}}$ \\
    \end{tabular}
  \end{ruledtabular}
\end{table}

Tab.~\ref{contributions} contains the values of the one- and two-loop
electroweak corrections in comparison with different components with a
single fermion loop for different values of the Higgs boson mass.  The
full two-loop result in the third column corresponds to the sum of the
fourth, fifth and sixth column plus the contributions with two fermion
loops as well as the effect induced by the running, $\Delta\alpha$,
of the fine structure constant. In the one-loop result we have
kept a finite $b$ quark mass, which has an impact of the order of
$-4.5\times 10^{-5}$. The perturbative expansion is performed in the
fine structure constant, $\alpha$, and not in $G_\mu$, since we want
to avoid any uncontrolled ``resummed'' terms.  The first observation
is that the third quark family contributions are very large, which is
expected, since they include the leading top-bottom mass splitting effects in
the $\rho$ parameter, $\Delta \rho$ \cite{Veltman:1977kh}. We have convinced
ourselves that the result has the correct behavior for large $m_t$
\cite{Fleischer:1993ub}. It is interesting that even though the light
fermion contributions in Tab.~\ref{contributions} do not contain the
running, $\Delta\alpha$, of the fine structure constant, they are
sizable. The last column gives the values of the tr$\gamma_5$
contribution. It has to vanish for vanishing mass splittings in the
fermion families  and can be at most logarithmic for large top quark
masses, which explains its smallness.   For small $M_H$, the total
two-loop result is rather small, but we note that this is due to a
fragile cancellation strongly dependent on $m_t$. With the older value
$m_t = 174.3$ GeV, the result would be of the order of $5 \times 10^{-4}$ for
all values of $M_H$.

\begin{table}
  \caption{\label{contributions} One-loop and fermionic two-loop
  electroweak contributions to $\Delta\kappa$ with $M_W$ as input
  parameter. The subscripts ``tb'', ``lf'' and ``tr$\gamma_5$''
  correspond to the contributions of single loops of the third quark
  family, of the light fermions (without the running
  of the fine structure constant) and of the
  ``tr$\gamma_5$'' effects in the triangle fermion subloops (see text).}
  \begin{ruledtabular}
    \begin{tabular}{cccccc}
      $M_H$ & ${\cal O}(\alpha)$ & ${\cal O}(\alpha^2)_{\mbox{\footnotesize ferm}}$ & 
      ${\cal O}(\alpha^2)_{\mbox{\footnotesize tb}}$ & ${\cal O}(\alpha^2)_{\mbox{\footnotesize lf}}$ & 
      ${\cal O}(\alpha^2)_{\mbox{\footnotesize tr$\gamma_5$}}$ \\
      $\mbox{[GeV]}$ & $\times \; 10^{-4}$ & $\times \; 10^{-4}$ & $\times \; 10^{-4}$ & $\times \; 10^{-4}$ & $\times \; 10^{-4}$ \\
      \hline
      100 & 438.94 & -0.63 & -16.96 & -2.84 & 0.27 \\
      200 & 419.60 & -2.16 & -17.10 & -3.08 & 0.27 \\
      600 & 379.56 & -5.01 & -16.89 & -3.77 & 0.27 \\
      1000& 358.62 & -4.73 & -14.90 & -4.25 & 0.27 \\
    \end{tabular}
  \end{ruledtabular}
\end{table}

In order to provide the most precise prediction for \SinEff in the SM we must
use the muon decay constant, $G_\mu$, as input parameter. The procedure to
derive $M_W$ from $G_\mu$ is described in detail in \cite{mw}. In analogy to
that work, we do not want to perform any ``resummations''. Instead, we include
both in $\Delta r$ and \SinEff all known effects in expanded form.  Besides the
electroweak two-loop terms presented above, these effects encompass QCD
corrections to the one-loop prediction at the two- \cite{QCD2L} and three-loop
level \cite{QCD3L} and also the recently obtained ${\cal O}(\alpha^2 \alpha_s
m_t^4)$ and ${\cal O}(\alpha^3 m_t^6)$ corrections to $\Delta \rho$
\cite{faisst}. We kept again a finite $b$ quark mass in the ${\cal O}(\alpha
\alpha_s)$ correction, which has an impact of $4.5\times 10^{-5}$, almost
completely canceling the similar effect in the ${\cal O}(\alpha)$ prediction.
Consistency requires that we also take leading reducible effects at ${\cal
O}(\alpha^2 \alpha_s)$ and ${\cal O}(\alpha^3)$ into account. It turns out that
separate terms as {\it e.g.} $c_W^2/s_W^2 \Delta \rho \Delta \alpha^2 $ are
quite sizable, but when summed cancel each other as seen in Tab.~\ref{qcd}. 
We stress once
more at this point that the same effects have been included in $\Delta
r$ and in \SinEff. This means in particular that, contrary to
\cite{mw}, we do not take the bosonic corrections \cite{bosonic} to
$\Delta r$ into account. Such precautions are enforced by the
sensitivity of \SinEff to $M_W$, since a 1 MeV shift in the latter
causes a shift of about $-2\times 10^{-5}$ in the former.

\begin{table}
  \caption{\label{qcd}
  Various QCD corrections to $\Delta\kappa$ and the only known pure three-loop
  electroweak irreducible contribution, stemming from $\Delta \rho$, 
  in comparison with three-loop reducible effects. The input parameter
  is $M_W$.}
  \begin{ruledtabular}
    \begin{tabular}{cccccc}
      $M_H$ & ${\cal O}(\alpha \alpha_s)$ & 
      ${\cal O}(\alpha \alpha_s^2)$ & 
      ${\cal O}(\alpha^2 \alpha_s m_t^4)$ & 
      ${\cal O}(\alpha^3 m_t^6)$ & 
      reducible \\
      $\mbox{[GeV]}$ & $\times \; 10^{-4}$ & $\times \; 10^{-4}$ &
      $\times \; 10^{-4}$ & $\times \; 10^{-4}$ & $\times \; 10^{-4}$ \\
      \hline
      100 & -36.83 & -7.32 & 1.25 & 0.17 & 0.92 \\
      200 & -36.83 & -7.32 & 2.08 & 0.09 & 0.94 \\
      600 & -36.83 & -7.32 & 4.07 & 0.07 & 0.97 \\
      1000& -36.83 & -7.32 & 5.01 & 0.99 & 0.98 \\
    \end{tabular}
  \end{ruledtabular}
\end{table}

Our complete result is summarized by the following fitting formula,
which  reproduces the exact calculation with maximal and average
deviations of $4.5\times10^{-6}$ and $1.2\times 10^{-6}$,
respectively, as long as the input parameters stay within their
$2\sigma$ ranges and the Higgs boson mass in the range 10 GeV $\leq M_H \leq$ 1
TeV,
\begin{eqnarray}
\label{formula}
\sineff &=& s_0 + d_1 L_H + d_2  L_H^2 + d_3  L_H^4 + d_4  (\Delta_H^2 -1) \nonumber \\
&& + d_5  \Delta_\alpha + d_6  \Delta_t + d_7  \Delta_t^2 + d_8  \Delta_t  (\Delta_H -1) 
\nonumber \\ && + d_9  \Delta_{\alpha_s} + d_{10} \Delta_Z,
\end{eqnarray}
with
\begin{eqnarray}
&& L_H = \log\left(\frac{M_H}{100 \mbox{ GeV}}\right),\;\;\;\;
\Delta_H = \frac{M_H}{100 \mbox{ GeV}}, \\
&& \Delta_\alpha = \frac{\Delta \alpha}{0.05907}-1,\;\;\;\;
\Delta_t = \left(\frac{m_t}{178.0 \mbox{ GeV}}\right)^2 -1, \nonumber \\
&& \Delta_{\alpha_s} = \frac{\alpha_s(M_Z)}{0.117}-1,\;\;\;\;
\Delta_Z = \frac{M_Z}{91.1876 \mbox{ GeV}} -1, \nonumber
\end{eqnarray}
and
\begin{equation}
\begin{array}{lll}
s_0 = 0.2312527, \\
d_1 = 4.729 \times 10^{-4}, &
d_2 = 2.07 \times 10^{-5}, \\
d_3 = 3.85 \times 10^{-6}, &
d_4 = -1.85 \times 10^{-6}, \\
d_5 = 0.0207, &
d_6 = -0.002851, \\
d_7 = 1.82 \times 10^{-4}, &
d_8 = -9.74 \times 10^{-6}, \\
d_9 = 3.98 \times 10^{-4}, &
d_{10} = -0.655.
\end{array}
\end{equation}
The impact of this result is shown in Tab.~\ref{difference}, where we
compare our prediction with the previous result as given in the
fitting formula in \cite{Degrassi:1997iy} and implemented in ZFITTER
\cite{zfitter}. The difference varies from roughly $-4\times 10^{-5}$
to $-8\times 10^{-5}$ for the $M_H$ range from 100 to 300 GeV, which
is the preferred mass region inferred from precision electroweak data.
These values reach half of the experimental error and induce an
important shift in the central value of $M_H$ derived from \SinEff alone.
With the most recent value of the top quark mass given in Tab.~\ref{input} the result
shifts the central value from 149 GeV to 168 GeV, to be compared with
the shift induced by the new $m_t$ measurement which gives a jump from
132 GeV to 168 GeV.
\begin{table}
  \caption{\label{difference} Difference between the result
    Eq.~(\ref{formula}) and the previous result including terms of
    ${\cal O}(\alpha^2 m_t^2)$ from \cite{Degrassi:1996ps}, obtained
    from the ZFITTER implementation (left column) or from the fitting
    formula from \cite{Degrassi:1997iy}.}
  \begin{ruledtabular}
    \begin{tabular}{ccc}
      $M_H$ & 
      $\left(\Delta\sineff\right)_{\mbox{\tiny ZFITTER}}$ & 
      $\left(\Delta\sineff\right)_{\mbox{\footnotesize \cite{Degrassi:1997iy}}}$ \\
      GeV & $\times 10^{-4}$ & $\times 10^{-4}$\\
      \hline
      100  & -0.45 & -0.40 \\
      200  & -0.69 & -0.72 \\
      300  & -0.85 & -0.83 \\
      600  & -1.17 & -0.94 \\
      1000 & -1.60 & -1.28 \\
    \end{tabular}
  \end{ruledtabular}
\end{table}
The formula Eq.~(\ref{formula}) has been implemented in the most
recent version of ZFITTER, version 6.40
\cite{zfit640}.

Besides providing an up to date fitting formula, it is necessary to
\pagebreak[0] discuss the error on the theoretical prediction
connected with the unknown higher order contributions.  Here one has
to incorporate the treatment of the error of the $M_W$ prediction,
since the final prediction for \SinEff takes $G_\mu$ as input. In
particular, there are some cancellations between the radiative
corrections to $M_W$ and the $Z$ decay form factors that go into
\SinEff.  We take the point of view that these cancellations are
natural and discuss both quantities in conjunction. To this end, we
use geometric progression from lower orders to estimate the missing
higher-order contributions and add them quadratically at the end. In
units of $10^{-5}$ we assign the following errors: corrections of
${\cal O}(\alpha^2 \alpha_s)$ beyond $m_t^4$ vary between 2.3 and 2.0
for $M_H$ between 10 GeV and 1 TeV, corrections of ${\cal
O}(\alpha^3)$ between 1.8 and 2.5, ${\cal O}(\alpha \alpha_s^3)$
between 1.1 and 1.0, ${\cal O}(\alpha^2 \alpha_s^2)$ between 1.7 and
2.4, and finally the bosonic corrections at ${\cal O}(\alpha^2)$ are
expected to be of the order of 1.2. This gives an error varying
between 3.3 and 3.5, and we take as our estimate the latter,
largest, value. To account for possible deviations from the
  geometric progression of the perturbation series, we included an
  additional overall factor of $\sqrt{2}$, giving a final error of
  $4.9\times 10^{-5}$.

\begin{figure}
\hspace{0cm} \epsfig{file=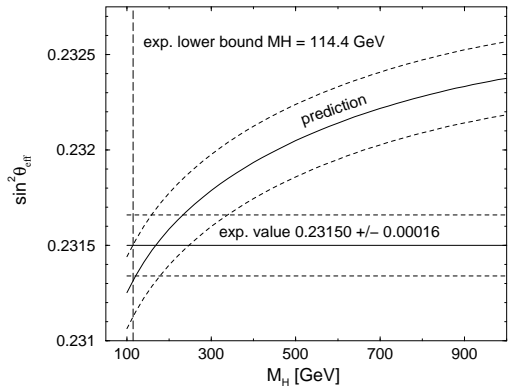,angle=270,width=6.6cm}
\caption{\label{experimentplot}
The \SinEff prediction against the current experimental value,
with 1$\sigma$ bands from the experimental input.}
\end{figure}

If we take into account all of the input parameter errors, our
prediction can be compared with the experimental value as
shown in Fig.~\ref{experimentplot}.

In conclusion, we have calculated the complete fermionic corrections
to \SinEff at the two-loop level and obtained a sizable contribution
when compared to the previously known leading and subleading terms in
the top quark mass expansion. Together with our result for the $W$ boson
mass and recently obtained three-loop terms, we are able to give the
most up to date prediction to be used in the global fit to electroweak
data. Furthermore, we implemented the result in the program ZFITTER,
widely used for this purpose.

The authors would like to thank P.~Gambino, F.~Jegerlehner and
A.~Sirlin for  useful comments.  This work was supported in part by
TMR, European Community Human Potential Programme under contracts
HPRN-CT-2002-00311 (EURIDICE), HPRN-CT-2000-00149 (Physics at
Colliders), by Deutsche Forschungsgemeinschaft under contract SFB/TR
9--03, and by the Polish State Committee for Scientific Research (KBN)
under contract No. 2P03B01025.

\end{document}